\documentstyle[12pt]{article}
\input prepictex \input pictex \input postpictex
\renewcommand{\baselinestretch}{1.2}

\author{Hans - J\"urgen Schmidt}
\title{Answer to question \# 55}
\date{}
\begin{document}
\maketitle

\centerline{
Universit\"at  Potsdam, Institut f. 
Mathematik}
\centerline{
      D-14415 POTSDAM, PF 601553, Am Neuen Palais 10, Germany}

\renewcommand{\baselinestretch}{1.2}

{\large {\bf
\noindent  
Answer to question \# 55 [ ''Are there pictorial examples 
that distinguish covariant and contravariant vectors 
?''
 D. Neuenschwander, Am. J. Phys. 65 (1), 11 (1997)]
}}

\medskip

 American J. of Physics in print 

\medskip

%\bigskip

Neuenschwander $^1$ asked how to visualize the
distinction between co- and contravariant vectors. 
 Most of all textbooks introduce this distinction
on an abstract level, the only exception I know is
Stephani $^2$, and below I will show how I present it
in my lectures ''Introduction to differential geometry'' 
at Potsdam university.

\medskip

If {\it no metric exists } at all, then covariant vectors and
contravariant vectors are different types of objects.

\medskip

If {\it a metric exists}, then there is a canonical isomorphism
between them; so we introduce {\it vectors}, and  after 
fixing a coordinate system, we speak about their covariant and 
their contravariant components. 

\medskip

In the following, we will deal with the second case only, because
it is more easy to visualize: The chalkboard has a 
canonical metric which makes it a flat two-dimensional Riemannian
manifold. 

\medskip

Neuenschwander $^1$ wrote 
 that the mentioned distinction is necessary when 
dealing with curved spaces. This is not wrong, but it is a little
bit misleading, and I prefer to say: ''\dots 
is necessary when dealing with a non-rectangular coordinate
system.'' Example: We fix a point (the ''origin'' $O$) in the
  Euclidean plane, then there is a one-to-one 
correspondence between points and vectors. (The point $P$
is related to the vector $\overline{OP}$ .) 
First we use  rectangular coordinates. We might
call them $x$ an $y$;  
however, as we are interested to see,
how the situation changes by introducing 
non-rectangular ones,
we call them $x^i$ with  $i \in \{1,2 \}$ .
% The $2$ coordinates will be denoted by $x^i$. 
So the point $P$ has coordinates $(x^1, \, x^2)$, cf. Fig. 1.

\bigskip 

\bigskip

\beginpicture
\setcoordinatesystem units <1cm,1cm> point at 0 0
\setplotarea x from 0 to 4, y from -1 to 3
%\put {\line(1,0){4}} at 0 0
%\circulararc 286.5 degrees from 1.5 0 center at 3 -2
%\put {\line(0,1){4}} at 0 0
%\put {\line(1,1){1.414}} [Bl] at 0 0
%\put {\line(-1,-1){1.414}} [Bl] at 0 0
%\put {\line(1,-1){1.414}} [Bl] at 0 0
\thicklines 
\put {\vector(1,0) {4}} [Bl] at 0 0
\put {\vector(0,1) {4}} [Bl] at 0 0
\put {\line(1,0){3}}  [Bl] at 0 2
\put {\line(0,1){2}}  [Bl]  at 3 0
\put {\large{P}} at 3.2 2.5
\put {\large{$x^1$}} at 3 -0.3
\put {\large{$x^2$}} at -0.3 2.1
\put {Figure \large{1}} at 2 -1.5
%%\grn3\abb9.tex 
\endpicture

\bigskip

\bigskip 

\bigskip

The coordinate system is a rectangular one, and so 
the component $x^1$ can be equivalently described 
as the perpendicular projection to the $x^1$-axis
 or as projection parallel to the $x^2$-axis. 

\bigskip

Let us now consider the case of an inclined system. 
Let the angle between the axes be $\alpha$ with
$0 < \alpha < \pi$.

\newpage

%\bigskip
%\bigskip 

\beginpicture
\setcoordinatesystem units <1cm,1cm> point at 0 0
\setplotarea x from 0 to 4, y from -1 to 3
%\put {\line(1,0){4}} at 0 0
\circulararc 76.2 degrees from 1.2 0 center at 0 0
%\put {\line(0,1){4}} at 0 0
%\put {\line(1,1){1.414}} [Bl] at 0 0
%\put {\line(-1,-1){1.414}} [Bl] at 0 0
%\put {\line(1,-1){1.414}} [Bl] at 0 0
\thicklines 
\put {\vector(1,0) {6}} [Bl] at 0 0
\put {\vector(1,4) {1.6}} [Bl] at 0 0
\put {\line(1,4){1.28}}  [Bl] at 1.7 0
\put {\line(0,1){5.15}}  [Bl]  at 3 0
\put {\large{P}} at 3.3 5
\put {\large{$x^1$}} at 1.7 -0.3
\put {\large{$x_1$}} at 3 -0.45
\put {\large{$x^2$}} at 0.5 4.5
\put { $\alpha$} at 0.4  0.4
\put {Figure \large{2}} at 2 -1.8

\endpicture

\bigskip

\bigskip

$x^1$ is the projection 
parallel to the $x^2$-axis, and $x_1$  is 
the perpendicular projection to the $x^1$-axis. 
We get  $x_1 = x^1 \, + \, x^2 \cos \alpha $,
i.e., 
 $x_1 = x^1$ if and only if $\alpha = \pi/2$. 
In general we get the following linear relation
$$ x_i \  = \  g_{ij} \ x^j$$
by the use of the metric $g_{ij}$, 
where $g_{12} = g_{21} = \cos \alpha$,  
 $g_{11} = g_{22} = 1$, and summation over $j \in \{1,2\}$
is automatically assumed. 

\bigskip

{\large {\bf References}} 

\bigskip

\noindent
[1] D. Neuenschwander, Am. J. Phys. 65 (1), 11 (1997).

\bigskip

\noindent
[2] H. Stephani, General Relativity, Cambridge University 
Press 2nd edition 1990, page 26. 
(In the first German edition, which appeared in Berlin in 1977,
it is page 35.)

\end{document}